\documentclass[12pt]{article}

\usepackage[utf8]{inputenc}
\usepackage{amsmath, amssymb, amsthm, mathtools, bbm, graphicx, fullpage, caption, subcaption, multirow, url, authblk, hyperref}
\usepackage[dvipsnames]{xcolor}
\usepackage[normalem]{ulem}
\usepackage{setspace}
\newcommand{\blkfootnote}[1]{\begingroup\renewcommand\thefootnote{}\footnote{#1}\addtocounter{footnote}{-1}\endgroup}

\title{The Future AI in Healthcare: A Tsunami of False Alarms or a Product of Experts?}

\author[1,2]{\small{Gari D. Clifford, DPhil}}
\affil[1]{\small{Department of Biomedical Informatics, Emory University}}
\affil[2]{\small{Department of Biomedical Engineering, Georgia Institute of Technology \& Emory University}}

\date{18 July 2020}

\begin{document}

\maketitle

\blkfootnote{This work was funded by the Gordon and Betty Moore Foundation, by the National Science Foundation under award number 1822378 (Leveraging Heterogeneous Data Across International Borders in a Privacy Preserving Manner for Clinical Deep Learning) and the National Institutes of Health-sponsored Research Resource for Complex Physiologic Signals (www.physionet.org) (R01GM104987). The content of this article is solely the responsibility of the author and does not necessarily represent the official views of the National Institutes of Health. The author declares no conflict of interest.

For information regarding this article, please contact the author via email. Address: Department of Biomedical Informatics, Emory University, Woodruff Memorial Research Building, 101 Woodruff Circle, 4th Floor East, Atlanta, GA 30322, USA. Phone: (404)-727-4631. Email: gari@gtech.edu
}

%
%

\newpage

\begin{abstract}

Over the last five to ten years, the significant increases in affordable and accessible computational power and data storage have enabled AI (or more specifically, machine learning) to provide almost unbelievable classification and prediction performances compared to well-trained humans. Most breakthroughs have been in somewhat artificial, relatively low dimensional, and time-invariant problems, such as classic board games ({\it e.g., Go, Chess, Breakout}), or more practically and notably, in self-driving cars. However, there have been some promising results in the much more complex healthcare landscape, particularly in imaging. This promise has led some individuals to leap to the conclusion that we will solve an ever-increasing number of problems in human health and medicine by applying `artificial intelligence' to `big (medical) data'. In particular, the highly incentivized move to electronic medical records over this same period, particularly in the United States of America, has led to much excitement in the domain of classification and prediction tools in clinical settings. The scientific literature has been inundated with algorithms, trained on retrospective databases, claiming to predict a spectrum of problems and events, from stroke to sepsis, readmission or death. Unfortunately, I argue that most, if not all of these publications or commercial algorithms make several fundamental errors.

This editorial, written as a companion piece to a keynote address I gave at the Society for Critical Care Medicine\footnote{https://youtu.be/Tv9ADevax8M} in February of 2020, discusses these issues and proposes some potential solutions, particularly in the framework of public competitions (or `challenges').  Although I began writing it in the summer of 2019, it comprises a series of thoughts and connected mini-editorials I have written over the last ten years, but never published.
This article's main thesis is that everyone (and therefore every algorithm) has blind spots. Therefore, there are multiple `best' algorithms, each of which excels on different types of patients or in different contexts. This is because each team develops their models with a particular training set or bias, just as human experts perform differently on different types of patients. In turn, this is driven by their own training biases driven by their environment, mentors, and even language (if we are to believe Wittgenstein).
I then argue that the logical solution to developing a `best' algorithm is to vote many algorithms together, weighted by their overall performance, their independence from each other, and a set of features which define the context (i.e. the features that maximally discriminate between the situations when one algorithm outperforms another).
This approach not only provides a better performing classifier or predictor but provides confidence intervals so that a clinician can judge how to respond to an alert.
Moreover, I argue that a sufficient number of (mostly) independent algorithms that address the same problem can be generated through a large international competition/challenge, lasting many months.  I also discuss the key elements of such an event, particularly in the context of recent examples from the PhysioNet / Computing in Cardiology Challenges focused on sepsis and arrhythmias. Finally, I propose the introduction of the requirement for major grantees to run challenges in the final year of funding to maximize the value of research, and to select a new generation of grantees.


\end{abstract}

\section{Introduction}
\label{sec:introduction}

We live in exciting times for healthcare, standing on the edge of a fourth industrial revolution. Yet, in many ways, we are mired in the past, crippled by technical burden and antiquated ways of using technology.
The advent of embedded computers in medical practice, and to some extent, the internet and the mobile communication revolution, have made isolated impacts in closed-loop systems, diagnostics, and remote consultation. However, the mode of practice of medicine remains firmly hierarchical and rooted in traditional social constructs, such as the practice of `rounding,' which dates back to World War I and II \cite{united1962medical}, and was a necessary product of mass casualty situations. The oral transmission of information at each shift change or round, with the laying on of hands, can be likened to traditional `campfire songs,' used before writing to transmit useful cultural heuristics and memories.

Never-the-less, the recently articulated notion of the `unreasonable effectiveness' of machine learning (ML)\footnote{In 2014, Yan LeCun delivered a lecture with this title, borrowed from the title of Eugene Wigner's seminal article from 1960 `The Unreasonable Effectiveness of Mathematics in the Natural Sciences' \cite{Wigner1960}.} has created a bubble of excitement. (Although I use the eye-catching term Artificial Intelligence (AI) in the title, I will avoid its use in this article from hereon, since AI is an overlapping and somewhat larger field, which generally refers to technology that mimics how humans behave, including the process of learning, whereas the term ML generally refers to the field of {\it how} the learning is achieved (and is used in both AI and other prediction or classification tasks unrelated to AI). ML, and deep learning in particular, has been used to successfully attack a series of complex problems, ranging from image searches to speech recognition, partially delivering on the promise of neural networks, which disappointed everyone back in the 1990's. It's important to note that neural networks have been around for decades, and deep learning itself is not as novel as we might be led to believe by the media. Rather,  Moore's Law for computation and its analog in storage capacity, together with the development of massive multicore GPUs (designed initially for computer gaming), have combined to push us across an imperceptible threshold. Perhaps most notably embodied by the realization of self-driving cars, deep learning on relatively low cost, low energy GPU-based (edge computing) systems that combine to provide real-time navigation in life-critical situations. Despite the fact that high profile accidents have occurred, these are much lower than the frequency of mistakes made by humans
(3.1 fatalities per billion vehicle-km, compared to 7.3 per billion vehicle-km driven in the US in 2018
\cite{wikipediaSelfDrivingCarFatalities}). The potential is enormous when you consider that road traffic accidents are the largest global killer of people between 5 and 29 years old \cite{CDC2017}, costing an estimated 3\% of each nation's gross domestic product.

However, it is arguably in healthcare where the greatest opportunity for preventable errors exists.
Diagnostic errors contribute an estimated 40,000 to 80,000 lost lives per year in the US \cite{PinnacleCare2015,Slawomirski2017}.
These numbers are comparable to (and cost far more, at an estimated US\$750 billion than) the reported almost $\approx$170,000 unintentional injury deaths in the US, most of which are due to accidental poisoning ($\approx$65,000), motor vehicle accidents ($\approx$40,000) and falls ($\approx$36,000) \cite{WorldHealthOrganization2015,Slawomirski2017}.
Given the scale of these health burdens, reduction through better systems is a perennial call of health care professionals and thought leaders over the last decade or more \cite{Gawande2007,Balogh2016}.

By analogy with the self-driving car industry, there is great potential to make enormous progress in the (similarly) life-critical healthcare arena. In particular, the same technology that is releasing humans from the errors we cause by driving is likely to revolutionize medicine in a similar manner - by providing additional oversight during processes that demand sustained attention. These changes have increased the capacity of the user (or healthcare professional) to deal with other simultaneous tasks.
However, the barriers to success are different in healthcare, since the data and labels are far noisier, and the consequences of the predictions or classifications are less certain.

Although it is exciting to see that
the scientific literature has exploded with the application of ML to medical data, the vast majority of research (if not all) focuses on retrospectively trained ML algorithms.
For routine classification tasks, such as reading images, there is enormous potential, although the use of the algorithms outside of the population used to train and test the algorithm remains a barrier to use. Perhaps more worrying are the many works claiming to predict a spectrum of problems and events, from stroke to sepsis, to readmission and death.
Several common mistakes are present in almost all publications, and most probably in commercial algorithms\footnote{It is worth noting that I believe that these issues apply to commercial algorithms as well as open research, because commercial algorithms are subject to less rigorous scientific review/scrutiny than those that are published in the scientific literature.}.
Specifically, these errors include:
\begin{itemize}
\item Including the missingness of data in the models and other culture-dependent variables such as length of stay (thus encoding clinical behaviors rather than underlying physiology);
\item Training and testing on either a single database or across all databases as a single set (reducing the chances that the model will work on new databases/hospitals);
\item Not accounting for under-represented groups in the data;
\item Not taking into account the temporal nature of the data, and confusing classification with prediction. Training on a retrospective cohort to predict an event, such as sepsis, even if you account for the relative rate of sepsis per unit time, ignores the fact that an algorithm must monitor continuously, so the false alarm rate is likely to be extremely high;
\item Using the wrong (traditional) cost function(s) for classical information retrieval such as AUC, AUPrc, Precision, Recall, Accuracy, F1, etc. In reality, the cost function should reflect the clinical behavior that would result from a prediction or classification, with a continuous time-variant fuzzy cost function related to repeated testing or the relative clinical cost of downstream treatment.
\item Ignoring how humans may respond to the alert/prediction, assuming it would be a one-time decision, rather than a watch-and-wait approach, or a series of downstream decisions;
\item Providing a binary prediction, rather than a fuzzy membership, which more closely reflects human thinking, where you can be a member of more than one category;
\item Not accounting for the differences in noise in the observations or varying quality of labels;
\item Failing to provide confidence intervals in a prediction;
\item Assuming that there is one `best' algorithm.
\end{itemize}

The lattermost issue is perhaps the most under-appreciated in modern data science.
There are perhaps multiple cultural reasons for this. First, although many sporting tournaments are often team events,
there is a tendency for us to idolize the individual best performer, and credit the one who pushed the ball over the goal line, rather than the ones that orchestrated the complex build-up that created the unique opportunity. (Although it is interesting to observe the differences in US sport, where `assists' are given some partial credit.)
In addition, the mode of approving devices and software from commercial companies
could also drive this singular `best athlete' mindset. A company usually patents a specific method of approaching
a problem, rather than an array of specific methods, since the latter is harder to defend and implement.
(Patents are often written to {\it generalize} a given method, but this just obfuscates the method, rather than opens the door for multiple algorithms to be used together.)
The medical device or software approval process also then applies to a single approach. I do not recall having read a patent for an ensemble method in medicine, and it's hard to imagine how the FDA or other regulatory bodies would view such an approach.
In many ways, it's entirely natural to imagine our own way is the best. If those of us at the cutting edge weren't over-confident in our ability to develop world-leading research, we wouldn't have the confidence to try in the first place. In reality, the pay-off for the effort is pretty remote.
Never-the-less, if the decision is extremely important, the wise person polls many people they see as experts or trusted parties, and
aggregates the decision from these pieces of advice, with a weighting based on the perceived integrity or expertise of each opinion source, plus some contextual factors (or other biases). Why do we not then extend this approach to scientific decision making more often, and place it in a robust mathematical framework?

In this article, I argue that an ensemble of independent algorithms, or a {\em product of experts}, can work in a formal aggregation framework to guarantee a better algorithm than any single algorithm. Importantly, the ensemble algorithm deals with edge cases and minority classes in a far better way than a single generalized algorithm.
In addition, this type of approach also provides an ability to estimate the confidence in the prediction,
in very much the same way that we start to trust hurricane predictions when all the track lines start to converge and indicate landfall at the same location.
I also argue that independence is extremely difficult to achieve, particularly with the increasing ease with which we can download each other's code, and the proliferation of standard data science libraries such as TensorFlow, Caffe, PyTorch, and Scikit-learn. Ironically, it is the spread of open-source codebases and the culture of posting repeatable science through public code repositories, which has increased this tendency for much of the field of ML to manifest as minor variants of the same formulae.
This increasing homogeneity of research software isn't an argument for concealing scientific methods and discouraging public libraries. Still, it does indicate
 that we must build better ways of identifying when a piece of code is genuinely novel or independent and is adding value to the overall prediction or classification task.
In the following section, I describe a public challenge framework that could create such a tool. While the framework may be rather general, I will try to stick to examples from the field of ML in healthcare.

\section{Designing a challenge to encourage independent methods and code}

While this section focuses on the elements of a challenge, much of it can apply to research in general, particularly data set composition, label quality, and evaluation metrics. Of course, the primary consideration is the topic of the research.

\subsection{Choosing a topic}
In the field of healthcare, it is essential to have a well-defined target outcome that is meaningful. In that sense, it has to be something that leads to a change in action that is likely to cause an improvement in an individual.
For example, the 2017 Challenge focused on classifying 30-second single-lead electrocardiograms into one of
 three rhythms (normal, atrial fibrillation, or other), or as too noisy to classify. Although there are many
 more rhythms, it is difficult to differentiate them without more clinical leads. Never-the-less, diagnosis of atrial fibrillation alone is enough to set off a sequence of referrals for further testing. This structure makes the 2017 Challenge an
 example of a well-defined problem.
An example of a poorly defined topic can sometimes be the simplest
 and most obvious at first sight. A few years ago, a well-known ML-focused competition forum requested data from my institution to run a competition on mortality prediction. I advised against the idea because mortality is a poor treatment target. Everyone dies, but it's the timing and reason for death that is important, and potentially what you can do, if anything, to avoid it. Death before discharge from intensive care, discharge from hospital, or twenty-eight days later is arbitrary,
creating a false dichotomy. Edge cases (those that die at twenty-nine days, for example) then create significant confusion for any classification or prediction algorithm. Moreover, the utility of such
 an algorithm is therefore highly questionable (outside benchmarking the performance of institutions), since there is no specific intervention or treatment that
 can be prescribed, and it is not clear that predicting death early in a hospital stay could change the outcome even if the underlying cause of the prediction could be identified.
In contrast to this, the 2019 PhysioNet Challenge's target
was predicting sepsis six hours ahead of clinical suspicion, using the Sepsis 3 Criteria.
These criteria provide a clear set of clinical markers and predicting six hours before
 any clinician can identify the signs of sepsis provides an actionable window that can significantly
 change a patient's trajectory and affect outcomes.

\subsection{Data set composition}

The data are the most important part of any challenge. It is essential to have at least three V's of big data: {\it Volume, Variety} and {\it Veracity}.
For time-series data problems, {\it Velocity} is also important - that is, we must sample fast enough to avoid classic errors such as aliasing. This latter point is often ignored in medical data and can lead to `ghost' signals that mislead the data analyst.

\subsubsection{Data variety}
It is critical that the data sets must represent the population in an unbiased manner. It is tempting just to include all the data you can find.
Of course, this leads to a hidden sampling bias, driven by the manner in which the data were collected, or the access to healthcare of the sampled population.
In addition, it is important that multiple datasets are used to represent the variety of ways in which different systems or institutions collect medical data.
Both of these issues are examined in more detail later in section \ref{sect:TestData}.
Conversely, selecting data based on which patients have full rank (no missingness), or artificially enhancing the representation of certain subgroups can lead to misleading results if the wrong metric is chosen. This issue is addressed more deeply in section \ref{sect:scoring_metrics}.

\subsubsection{Data volume}
\begin{figure*}
\centering
\includegraphics[width=17cm]{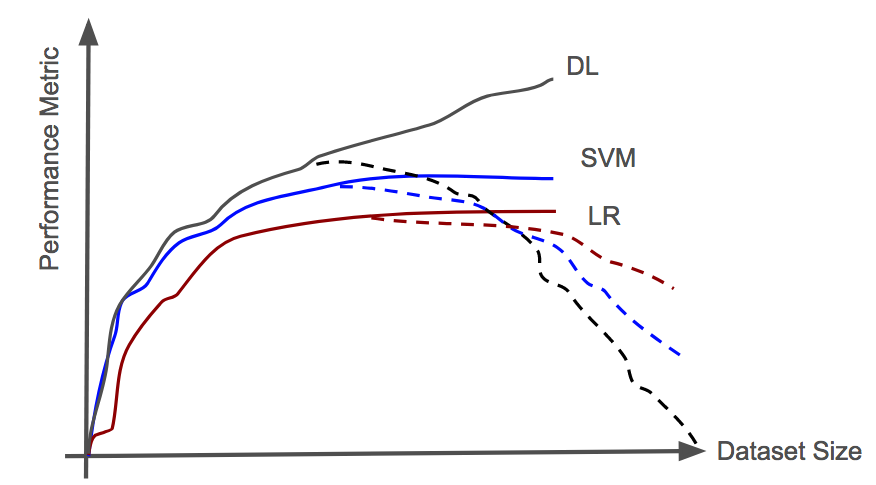}
\caption{The deep learning myth in medical applications. DL = deep learning. SVM = support vector machine. LR = logistic regression. 
Solid lines indicate how the algorithms should perform as dataset sizes increase. Dotted lines reflect the reality, because label quality drops as dataset size increases.
Figure adapted from \cite{CliffordCCMtalk2020} under the Creative Commons Attribution License 3.0 (CCAL). }\label{fig:MLmyth}
\end{figure*}

Data volume (quantity) depends on the question you are asking, the variety of the data, and the technique applied to the data to solve a given problem.
There is a meme in deep learning and big data, that as the data set size increases,  conventional ML approaches performances will level off
. Deep learning, on the other hand, is expected to improve performance as the dataset size increases. Assuming an infinite network topology and compute time, this may be true.
 (See solid lines in figure \ref{fig:MLmyth}.) However, just how big is big?
In the 2017 Challenge, we posted over 10,000 ECGs for a public dataset for a relatively simple four-class problem. Yet, the winning entry was a standard ML approach based on hand-crafted domain-expert-driven clinical features. Deep learning did not outperform this technique, and a well-publicized method by a Ph.D. student of a well-known Silicon Valley ML expert ranked behind some novice deep learners! This result demonstrates that ten thousand ECGs are not enough for deep learning to make an impact.
It does seem that one million ECGs may be enough to provide significant improvements through deep learning \cite{Attia2019}. However, this well-publicized study did not strictly compare deep learning to standard ML, or even multivariate regression for that matter.
While standard statistical methods are amenable to power calculations, it's hard to determine, {\it a priori}, just how big a dataset needs to be for ML approach to reach a given level of performance. The general approach, therefore, is to attempt to assemble as much data from as many subjects as time and funding permits. However, this in itself leads to another limitation, that of data quality or veracity.

\subsubsection{Data veracity - the emperor's new clothes}

The quality of data, and the associated labels are perhaps the most important, yet understated problems that ML faces in this domain.
Traditional databases, such as those found on PhysioNet prior to 2005, were relatively small by today's standards
but were meticulously hand-annotated. With the advent of large public datasets, it became impractical to have
expert (or even non-expert) annotation of all the significant events in the database. An example is the MIMIC II database,
comprising over 30,000 patients stays, each lasting multiple days, and for a significant subset, including hundreds of
thousands of hours of electrocardiographic and other bedside monitor data \cite{Saeed2011}. With monitors triggering
 events every few minutes on almost every patient, and up to 95\% of them being false \cite{Aboukhalil2008}, it is
impossible to determine the veracity of each event. Even when data is verified in real-time by the clinical
staff, this process can lead to significant errors \cite{Hug_CCM2011}.
There has been so much interest in the application of ML to medical databases, and EMRs in particular, in the hope that the complex associations of the data with specific events could be identified. However, this dream has been thwarted because there are vast limitations in these databases which inhibit their use. I often refer to EMRs as the `Emperor's new data', because it has been touted to be a gold mine of finery, but then during the weaving process, the fabric was left out, and we were left naked (as far as a useful predictive tool). Data collected for routine clinical activities are recorded for human review or billing.  EMRs were never designed for predictive analytics and ML. In the next section, I explore this fundamental issue, and in particular, the issue of errors in the labels.

\subsection{Label quality limitations and the big data challenge}

Label errors come in many forms in medical databases. These include missing labels, false or phantom events, incorrect class labels (because humans do not over-read the labels), or temporal misalignments. (In EMRs for example, the label can be provided hours after the event, increasing the likelihood of remembering the event incorrectly, and leading to a timestamp that is minutes or hours before/after the event, destroying causality in the data!) Noise can also creep in from differences in ontologies, protocols, and end-usage of the data. For example, when using billing codes as targets, it is important to realize that the codes are sometimes optimized for reimbursement, not medical accuracy.

Even in well-annotated databases, the over-read gold standard labels still contain significant noise and errors, and the target labels are usually given by humans. Moreover, the quality of the labels is bound to drop with increasing dataset size, as the humans have to cut corners, we use annotators with lower expertise, or we use algorithms trained on the smaller data sets. Never-the-less, it is important that the data set used has high inter-observer agreement levels. To improve agreement levels, multiple independent experts are required to label the data. In the 2017 PhysioNet Challenge \cite{clifford2017af}, we found that for a four-class (arrhythmia classification) problem, even eight experts were insufficient to reach a consensus for a significant portion of the labels.

In reality, there is an optimal data set size that a finite group of humans can curate, and beyond that, the algorithms must be trained on increasingly lower quality data. In fact, the highly nonlinear nature of more complex algorithms means that simpler algorithms may even be preferred in such situations.
(See dotted solid lines in figure \ref{fig:MLmyth}.) Michael I. Jordan, a well-known and respected ML expert at Berkeley, has been famed for
predicting the `Big Data Winter', or rather the failure of ML to live up to the hype. I believe that label quality will be one of the main factors that may lead to such a winter. When subsequently asked for his opinion on which ML algorithm he was most excited about Jordan was reported to have said, in all seriousness I believe,  `Logistic Regression'. We hear about `the unreasonable effectiveness of ML', but little is said about {\it the unreasonable effectiveness of logistic regression}, an algorithm with roots that stretch back to the early 19th century \cite{RePEc:tin:wpaper:20020119}. Notably, Claudia Perlich won several international competitions using this relatively simple approach. It is remarkable that this algorithm, which is essentially a simple neural network\footnote{with one hidden layer with a single hidden node, an identity activation function, and a single output node with the logistic sigmoid activation function}, or a discriminative form of na\"{i}ve Bayes \cite{NIPS2001_2020}, performs so well. One of the reasons for this is that logistic regression, particularly when combined with some form of penalization for overfitting, has far fewer free parameters than most classifiers, and is, therefore, less likely to overfit on noisy labels.

Recent developments in ML aimed at dealing with noisy labels hold promise \cite{NIPS2013_5073}, but have yet to demonstrate anything significant in medical data.
There is, however, another approach - that of {\it crowd sourcing}.
The `wisdom of crowds' asserts that given enough independent annotators, the average will tend to the right answer.
For most challenges we use a crowd of experts and vote together the labels to create a more accurate label, and a confidence interval (quantified by the spread of values or categories for the annotation of the same event). There are multiple issues with this approach, though. First, aggregating individuals, even low-cost individuals sourced through websites such as Amazon's Mechanical Turk, is not that scalable. Each annotation costs around \$0.1. With typically 20-50 annotations required to collapse the uncertainty of a label to acceptable levels, large databases can require millions of dollars to label accurately. Secondly, it's not clear how much one should weight each annotator, but equal weighting is unlikely to be optimal. A third related issue is that there may be multiple schools of thought about an annotation or label. If the distributions of annotations are highly skewed or multimodal, then the average might be completely wrong, and not reflective of any school of thought.  Finally, it is very hard to determine if annotations are truly independent.  In reality, annotators are rarely fully independent, with individuals having been educated in similar schools, or by the same expert. In some cases, experts often confer and reduce the information gain provided by voting. In section \ref{sect:crowdsource}, I  discuss how these issues can be addressed.

\subsection{How much preprocessing and cherry-picking should we do?}

In addition to label errors, we also expect noise and outliers in the source data.
The question of how much data to filter out or change before releasing them to the public is one of the most fundamental questions for any challenge. Ideally, the data would be provided in as raw of a representation as possible.
However, this sets up a large barrier to entry, and reduces the chances to accelerate innovation in certain areas. To address this tension,
we provide example source code - a baseline algorithm - which implements significant preprocessing and attempts to solve the problem, sometimes
with state of the art algorithms. In this way, users have the opportunity to significantly build on prior work.
The downside to this is that it may drive challengers to produce similar code bases, reducing independence.

When providing multiple databases, some level of normalization is needed to make sure that the differences in acquisition systems or their settings aren't learned as features. If one or more data sets have biases in the distribution of the classes of outcomes of interest, then it is possible that an algorithm, particularly a deep learning approach, would learn the nuances of the difference between the format of the data, and associate them with the differences in class distributions. For example, one dataset may be drawn from one hospital that deals with more severe cases and has a higher proportion of a given race. A deep learning algorithm might learn that this particular race is always sicker, and create a high false positive rate of treatment for this class \cite{Obermeyer447}.
On the other hand, it is sometimes important to preserve these differences, particularly in the unseen data, to encourage users to develop algorithms that are insensitive to these differences. Unfortunately,  even when two or more public databases are provided, challengers often develop algorithms over all the training data, and do not exploit the differences. Worse, they may add a flag that represents from which database the data are drawn, and thus ensure a lack of generalizability.

\subsection{Challenge length and phases}
\label{sect:challengelength}
The length of the public challenges ranges from 24 hours (the usual `hackathon' format) to several years (such as in the case of the X-prizes).
As noted already, to solve a significant problem in healthcare, a thoughtfully labeled data set (or set of data sets) is required.
It can take several years to assemble, format, label, and prepare for public dissemination. We often search for datasets years in advance.
However, even after the first release of a well-curated dataset, there needs to be a significant {\it Beta Test Phase},
in which the public can comment on the data and metric(s) proposed for the target.
Community feedback on the design of the challenge is an essential ingredient, allowing our peers to provide input, without
being prevented from participating. For the challenge, we usually run the Beta Test Phase for two to three months, to
allow us to stress test the entire framework for challengers to submit code, and the response of any performance metrics.

To incentivize challengers to participate in this stress-test of the challenge framework, a single successful entry during the Beta Test Phase is required to
be eligible for the final prize. Moreover, challenge teams are allowed to submit up to five entries, giving them more experience and time to develop
a better algorithm. Although any entry during this phase doesn't officially count towards their final score, it does provide insight, and
recently we showed that submitting several early entries is associated with a higher final ranking.

The challenge is closed for a week after the Beta Test Phase, and the organizers regroup to examine what rules, metrics, and data should change in response to feedback from the challengers. The Official Phase of the competition then runs for at least four, sometimes five months, providing challengers ample time to develop world-class algorithms.
In general, ten official entries are allowed, which are run on a subset of the hidden test data, and the scores are pushed to a public leader board within a day or two of submission.  In general, the subset is restricted to less than 30\% of the hidden test data to prevent sequential over-training on these data.
Approximately one month after the Official Phase opens, the teams are required to submit an abstract describing their preliminary approach to a conference at which they will present their approach. The abstract is reviewed for quality, and low quality ones are rejected. Since the publication of a scientific article is an important element in the challenge, this review step is essential.
After the closing date of the challenge, the teams are given the option to nominate their preferred algorithm out of the ones they have submitted, to be run on the full test data. (If no nomination is received, then the algorithm which produced the best score so far is selected.) This score is the only one that counts since it includes the full test data.

\subsection{Compute cost, time and memory limits}

In any public challenge, there have to be limits on the amount of time any algorithm takes to complete the task.
Setting the maximum compute time and memory capacity is challenging. One must factor in the cost of overall computation cost versus the downstream benefit, and the practicality of implementing the algorithm in real-time on affordable hardware. There is also a psychology of the software development team to consider. Most teams wish to react to code within 12-16 hours of submission, particularly if it failed to complete a run across all data. Therefore, we set an upper limit of 24 hours clock time on a well-provisioned cloud system. Each year this changes, but turns out to be equivalent to costing about \$0.1/hour, or much less than \$0.01 a patient/subject.

\subsection{Scoring metrics}
\label{sect:scoring_metrics}
The metric for a given challenge is always one of the hardest issues to debate. The choice, of course, depends on the data, the problem, and the downstream practical application.
We modify the challenge framework at least once, sometimes more during the course of the event. This sometimes includes our metrics. We do this because we acknowledge that our choice may not be `optimal', and we feel the scientific community should have a vote in this (within reason). Our aim isn't to run an artificial competition where thousands of teams beat the data to death to achieve the highest score on a standard metric, then claim they are the best team in the world. Rather, our aim is to push forward the frontiers of knowledge and research on a specific research topic, encouraging researchers to question everything about the structure of the challenge from the data to the scoring metric. (Indeed, one of the most important outputs of the challenges are the novel scoring metrics\footnote{The other key products of the challenges are the exchanges between the researchers, the resultant codebases, and subsequent scientific publications.}.) 

In the 2020 Challenge, focused on classifying arrhythmias from twelve-lead diagnostic electrocardiograms, we created yet another novel scoring metric. This metric treated any rhythm that would lead to the same (or very similar) treatment as almost the same. The result was a large confusion metric that represented the cost of classifying any rhythm as any other rhythm. 
In the 2019 Challenge, we developed a new scoring metric for predicting future events that accounted for the repeated nature of the prediction.  Metrics such as `area under the curve', recall, precision, and other conventional information-theoretic measures deal poorly with the relative rarity of the event being predicted, and the fact that there isn't just a single point in time at which it is useful to make the prediction. I.e., predicting sepsis 4 hours ahead of time is almost as useful as six hours ahead of time, and both predictions are much more useful than two hours ahead of the event.  

We have long debated whether we should include a metric for computational efficiency or cost in the competition, but this adds in another time-varying dimension to the metric. Does it matter if you spend \$ 10k on cloud compute or do it on a recycled mobile phone? Probably, but what if the two costs become negligible? Perhaps we should consider the environmental impact of the source of power that is used to generate the entry or the memory needed.  What if you can do it on a cheaper chipset - should GPUs be valued differently to traditional multicore processors? Perhaps there should there be an equity angle. For example, if more than 90\% of earth cannot afford or access the computing needed to do this, then should your entry be allowed?

The answer probably is: it depends on how the code is going to be used and what computing/memory costs in the future. A real-time vigilance detection system will need a low energy consumption / small memory footprint. An offline system for sleep staging for later review could use as much memory and compute power as the market will tolerate (e.g. compared with human labor, which is considerable in this case).

\begin{figure*}
\centering
\includegraphics[width=12cm]{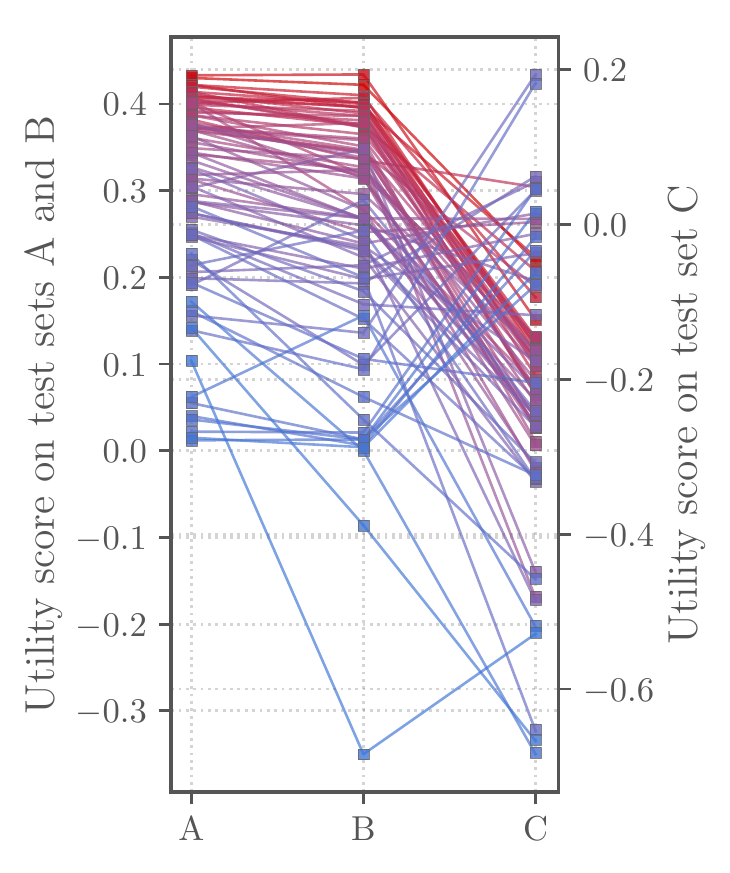}

\caption{\label{fig:PerformanceChallenge2019} Performance of top 70 competitors in the 2019 PhysioNet Challenge \cite{PhysioNet2019} on three hospital databases for sequestered (hidden) test data. Color coding indicates overall score across all databases (red indicating a higher performance). Only database C was completely hidden from the competitors. Lines join the same algorithm/team so that performance can be traced across datasets. Note that the scale for database C is different (right-hand side) and significantly lower. Note also that relative ranking and absolute score on the top-performing algorithms on databases A and B (and overall) did not correlate with performance on database C, indicating that no high scoring algorithm was able to generalize to a new hospital fully. Figure based on \cite{PhysioNet2019}, courtesy of Matt Reyna under the Creative Commons Attribution License 3.0 (CCAL).}
\end{figure*}

\begin{figure*}
\centering
\includegraphics[width=16.5cm]{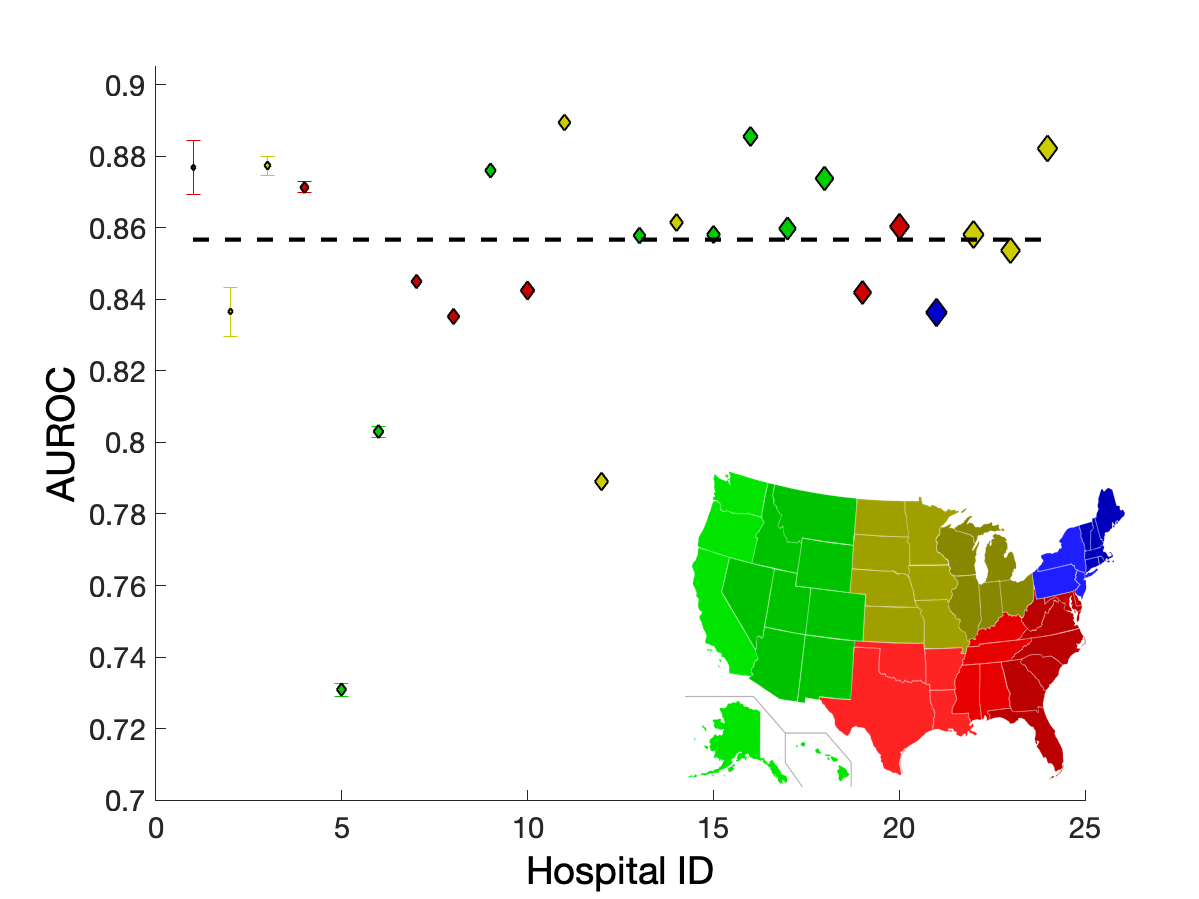}

\caption{\label{fig:LR_plot_Cerner} Cross-validation performance of a logistic regression model trained on 25 hospitals (200,000 patient stays) to predict mortality. Average AUROC indicated by black horizontal broken lines. Color codes indicate region in US per map, and size of diamond indicates relative contribution to model in terms of number of patients. Note that the performance is biased towards the West Coast and Mid West. Figure adapted from \cite{Clifford2011} under Creative Commons Attribution License 3.0 (CCAL).}
\end{figure*}

\subsection{Independent and hidden data - submitting code not labels}
\label{sect:TestData}

For each challenge we aim to provide at least three datasets of the same type of data, drawn from three independent sources.
This is important because it allows users to create algorithms that generalize across databases and can be tested on at least one database that is completely hidden from the user. This means that, in contrast to other competitions, we require the challengers to submit the code, not the labels. This prevents the challengers from training on the test data, or attempting to learn labels by repeated testing, unsupervised learning approaches, or even hand-labeling the data. The simplest way to cheat on the test data is to use the mean and variance to normalize the data algorithm. The test framework is constructed to prevent this so that challengers cannot access more than one individual at a time, and gaming of the test data in this manner is prohibited. Of course, the repeated testing allows the challengers to learn {\em some} non-specific information from the test data. This is consistent with the fact that we have observed that on repeated testing on hidden data, almost every entrant in the challenges has improved their score sequentially. For this reason, we only test on a small subset (typically 20\%) up to 10 times per entrant. When the challengers' best algorithms are run on the final data set, they observe an average of about a 5\% drop in performance.

\subsection{Generalization performance on multiple datasets}
It cannot be over-emphasized just how important it is to have multiple databases, and for one or more of those to be hidden. It is unreasonable to expect that an algorithm trained on one database would generalize beyond the center at which the data were collected. Johnson {\it et al} \cite{Johnson2018-DBLP:journals/corr/abs-1812-02275}
illustrate how poorly an algorithm trained on the MIMIC II database performs on the Philips VISICU eICU data from multiple other centers. This effect is also seen in the 2019 PhysioNet Challenge, where groups that performed well on the two public databases, tended to perform poorly on the hidden data from another medical center (Fig. \ref{fig:PerformanceChallenge2019}). This effect plays out in a non-random manner, based on the sampling bias. Figure \ref{fig:LR_plot_Cerner} illustrates how a logistic regression model for predicting mortality, trained on multiple hospitals using the Cerner ICU EMR, exhibit a wide range of performances, with the southeast of the US exhibiting particularly poor performance compared to hospitals in the western and mid-western areas of the US. This is illustrative of how important diversity is, yet on its own, it is not enough to solve the issues of generalizability beyond the training data.

\subsection{Publication and defense of solutions}
Many competitions take an easy way out, and post the unlabelled test data and require only the submission of the estimated labels or predictions. Without providing the code to actually implement the algorithm that produced the test labels, and a test to ensure the code actually runs and performs on a general system, any submission is of little value. Moreover, if the authors are not willing to defend their approach in a public forum (Computing in Cardiology),
then there is little incentive to describe the code and approach clearly.
Scores are emailed to challenge teams about a week after the end of the challenge, and teams must then prepare a four-page preprint describing the scientific approach, to be posted on the conference website for public viewing. The teams must then attend a public forum in person and orally defend the work.

Importantly, the final scores are not provided until after the challengers meet at the international conference
to discuss and defend their approaches at a public forum. This discussion is essential to the challenge, creating new knowledge, and enhancing the challengers' articles.
After the prize ceremony at the conference, challengers receive their final scores and are required to update their articles to reflect these scores.
A second peer review is then performed to ensure that each article is coherent and accurately reflects the final algorithm's performance.
It's always surprising how often authors will provide misleading results, using inappropriate metrics other than the official ones in the challenge,
or quoting training scores, or scores on a small subset of the test data. Ensuring that each team reports results that are directly comparable is
key to the scientific integrity of the challenges.

Although we allow teams to enter the challenge unofficially, and provide them with results, we always stress that such results should not be
 taken too seriously. Those teams that are unwilling or unable to fully describe their methods in peer-reviewed articles, provide little
 contribution to the field, and may have scored well by pure chance.

\subsection{Why go to a conference at all?}
While most of this article was written before the current pandemic, the pivot away from in-person meetings, particularly scientific conferences, has made me think even more deeply about whether we should even meet in person to defend our ideas. I'm on the board of Computing in Cardiology, so I've been privileged to witness and participate in the torment of having to rapidly move from in-person to online, and back again, depending on the ebb and flow of the pandemic and the risk-tolerances of everyone involved. While I'm not attending in-person this year, I do feel in-person attendance at conferences is important. It's almost impossible to understand if you trust an individual's research until you have an in-person meeting with them. It allows you to go beyond the biases you have, and get to know them as a person. Perhaps this creates new biases, because you may resonate with their love of a given sport, or have children the same age. This is something we struggle with constantly, often without an awareness of the issue. But it's very hard to use a video link to sit with someone for three hours over dinner and dive deeply into their thought processes. In the end, science is based on trust. We must trust our peers that they are honest and want to push the field to new heights, with an open mind to new ideas that are not their own. Opening up code and data helps, but the trust is bi-directional. We also need the challengers to trust that we are no manipulating the test data or scoring functions. For these reasons, we continue to require challengers to defend their approaches publicly.

\section{Other considerations for maximizing the utility of solutions}

\subsection{Code quality and independence}
\label{sect:Code_quality_and_independence}
The utility of the software submitted to data science competitions depends on four main factors:
\begin{itemize}
\item Reusability: To ensure a codebase entered into a competition is fully reusable, it is important that it can be run on an entirely new system from the one on which it was developed. By containerizing the pipeline, we reduce the probability that the code will fail, but it does not remove it completely since even small differences in random number generators can lead to differences in the training or even forward predictions of a complex model.
\item Documentation: Without good documentation concerning how the code was constructed, and importantly why the code was designed the way it was (justifying choices of coefficients and hyperparameters), it is very hard to trust any codebase or its authors.
\item Generalizability: By keeping a dataset completely hidden from public access, we are able to test how well any code generalizes to data from an unknown source.
\item Independence: This is perhaps the most overlooked issue. In section \ref{sect:crowdsource2} we discuss the utility of combining together large numbers of algorithms in a rigorous weighted framework to leverage the strengths of individual algorithms in different contexts. Although there has been much work on combining multiple labels on data from independent labelers (which could be humans or algorithms), almost all research has focused on combining independent annotators. When the annotators (or algorithms that generate the labels) are not independent, then there is no guarantee that an optimal combination can be found. Rather, incorrect, but similarly behaving algorithms, can reinforce each other and lead to a biased or incorrect aggregate label or prediction.

\end{itemize}

\subsection{Prize money - does size matter?}

Data science competition prizes have ranged from millions of dollars to nothing (i.e. kudos/bragging rights). The question of the size of the purse is entirely open. Obviously, the larger the amount, the more teams are likely to enter, but this doesn't guarantee a linear quality increase or a linear increase in independence between codebases (see section \ref{sect:Code_quality_and_independence}).

At PhysioNet, we never reveal the actual dollar amount of the prize. This is partially because the amount changes from year to year depending on the sponsor, but also because we feel that this is beside the point. The driving motivation to enter the Challenge should be the desire to solve problem itself, although we understand that humans are often motivated by more than one factor. This issue is discussed in more detail, together with why someone might want to {\em run} a challenge, in section \ref{sect:incentives}.

\subsection{Software and data licenses}

Several authors have discussed the benefits of crowdsourced code from competitions
\cite{Bender2016, Ledford2017, Guinney2018}.
For the PhysioNet Challenges, we encourage teams to use an open-source license, so that others may use the work for research or even translation/commercialization. It is notable that not all open source licenses prevent commercialization by third parties. One such example is the Berkeley Source Distribution (BSD) License. Moreover, this type of license does not prevent the author from patenting the approach and licensing it out to a commercial partner. Of course, it does not stop an industrial competitor from stealing the idea, but then so does publication in a journal or as a patent (in theory). Patents were never intended to create intellectual `property,' and bestow `ownership' of an idea to a particular entity. Rather, they were intended to provide a limited reward for contributing the knowledge into the public domain on how to reproduce the `invention.' The reward is limited to a decade or two without competition, giving the licensee a commercialization head start.
Of course, industry generally feels uncomfortable with this, and (almost) all patents in the computational field are as far from a scientific paper enabling reproducibility as possible, making general statements, sweeping claims and omitting important details (such as the exact parameters used, the precise model architecture, the details of the preprocessing steps and the exact values of the thresholds, coefficients, and weights in the model). However, it is not just patents that suffer from these ills. Many scientific papers also do, although not in such an extreme manner. Any complex piece of software cannot be properly described in an eight- to twelve-page paper (the limits most journals require), and so important details are edited out. It is also true that an author will often assume some concepts are obvious or trivial (which they may be to them at least), but to almost everyone else, they are not. It is for these particular reasons that we require all code to be submitted for testing on our servers (or rather on our cloud providers' servers), rather than providing unlabelled test data and asking for the challengers to submit labels. In this way, we know the code works in an environment beyond the developer's machine (and doesn't contain hardcoded paths, hidden files, and libraries, etc.), and the public can inspect the code to work out {\it exactly} how the authors actually designed their code.

Nevertheless, it is important that industry participates in such challenges, so that the challenge cannot be accused of being an academic exercise, and so that the winners reflect the best approaches in the field. We hope that the challenge data serve as a benchmark for a given problem or field, and that industry uses it for commercial testing and reporting.

It is worth noting that there are several public competitions that impose back doors to designers of the competition that mostly benefit them.
While one may argue that there should be some reward for the enormous effort it takes to design and run such a competition, the intentions of the organizers should be front and center.
Notably, Kaggle hosts competitions for third parties, which offer data and prize money in exchange for ``a worldwide, perpetual, irrevocable and royalty-free license ... to use the winning entry'', assigned exclusively to the third party sponsor of the competition \cite{KAggleTermsofUse2019}.
While this can encourage third parties to provide large datasets and significant prize money, attracting thousands of competitors, this may lead to a skewing of the results towards those hunting for large sums of money. (See the section on prizes for more discussion on this.)
Some competitions (e.g., Orange Data for Development cellular networks competitions \cite{Blondel2012,DeMontjoye2014}) only make the data available for a limited period of time and require the user to request permission to perform research on the data every time they want to use the data after the competition. This barrier significantly devalues the utility of the public data, since there is an arbitrary bottleneck for future use. Open datasets, without restriction, can provide benchmarks for the field, and help spur innovation on into the future. The original MIT-BIH Arrhythmia Database on PhysioNet has been around for almost thirty years, and although relatively small by modern standards, is still used as a benchmark database in FDA filings and many publications.

Perhaps one of the more open competitions (in terms of licensing and publications) is the {\em Dialogue for Reverse Engineering Assessments and Methods (DREAM)} Challenges, which are run in collaboration with Sage Bionetworks. The organizers state that the objective is to run `Community competitions on fundamental questions about systems biology and translational medicine, and advance computational methods.'
When submitting model code, participants should provide it under an
``open-source license of their choice'' which must ``permit the DREAM Challenges and Sage Bionetworks to distribute the code to the public for noncommercial research and development use via Synapse''. Authors are allowed to retain copyright and, of course, patent the idea before submission \cite{DREAMChallengesFAQs,DREAMChallengesAbout}.
\cite{Bender2016, Ledford2017, Guinney2018}.

\subsection{Repeatable code - requiring the training code?}
While there is an increasing tendency for journals to request open-source code (and more commonly, open-access data), evaluating code for replication of results is a non-trivial task. Only the diligent reviewer will thoroughly vet the code. However, I have never seen a journal or competition demand the authors submit code to train the model. Without this step, the research will never be truly replicable. For this reason, in the 2020 PhysioNet Challenge, we have required the users to submit the training code as well as the forward model.

\subsection{Ethical computing}
With the rise of big data and ML, we are confronted by two critical issues. These are 1) the environmental impact, and 2) the potential for bias to create algorithms that disproportionately disempower minorities and disadvantaged populations. 

The impact of the energy consumption required for large scale machine learning has recently attracted attention, with estimates 
that data centers could account for 10\% of total electricity consumption by 2025 \cite{Andrae2015,Andrae2017}.
It has been often quoted, that training a single AI model can emit as much carbon as the total lifetime of five cars \cite{DBLP:journals/corr/abs-1906-02243}! (Specifically, a Transformer consisting of 213 million parameters, with a neural architecture search.) 

Of course, promising developments in machine learning are reducing the computational complexity of ML algorithms, often with little impact on performance. Frankle and Carbin recently showed that neural networks contain subnetworks that are up to one-tenth the size of the overall system, yet are capable of being trained to make the same predictions without loss of performance, and sometimes can learn to do so even faster than the original much larger network \cite{DBLP:journals/corr/abs-1803-03635}.

While we restrict both the training (and separately) the test computation to 24 hours on state-of-the-art equipment\footnote{In 2020 we ran the training code on Google Cloud using 8 vCPUs, 54 GB RAM, and an optional NVIDIA T4 Tensor Core GPU and the trained model on Google Cloud using 2 vCPUs, 13 GB RAM, and an optional NVIDIA T4 Tensor Core GPU.}.
We have debated whether future Challenges should restrict this compute performance to lower performance edge computing systems such as the Coral TPU running Tensorflow Lite. However, this may distract from the main task of solving the medical problem first. The task of converting it to a more efficient model usually follows. With increasingly larger carbon footprints, this paradigm may need to change, however.



The concept of bias in ML is perhaps more tricky, but none-the-less, not less important. Multiple researchers have raised the issue of race and gender bias in AI \cite{ONeil2016,pmlr-v81-buolamwini18a,10.1145/3306618.3314244, Obermeyer447}. Several have even proposed frameworks to measure and deal with this issue \cite{DBLP:journals/corr/abs-1710-03184,DBLP:journals/corr/abs-1710-06876, NIPS2017_6988, DBLP:journals/corr/abs-1808-00089,Sahil2018,Bellamy2019}.  
These issues, and frameworks, should be increasingly addressed in public competitions, both in the scoring systems, and the literature exploring the results of each event.
In section \ref{sect:crowdsource2}, I discuss ways to reduce bias from multiple biased algorithms. I also argue that, because we cannot escape our intrinsic biases, public challenges, which generate a spectrum of biases, may indeed be the best way to mitigate the effect of these biases. 

Future challenges will explicitly incorporate these issues into the scoring metrics whenever possible. 
For example, in the 2020 challenge, the race and gender of many of the subjects is known. A post-challenge assessment will identify how algorithms perform across race. In later challenges, this information could be used to explicitly address bias, and perhaps become part of the scoring metric, penalizing those that show unbalanced performance across race and gender. It is unclear how we might offer a trade-off between performance, bias, and efficiency. Therefore, it may make sense to run three challenge categories, constrained by having to be in the top 10 (say) of each category.

\section{A commons of models?}

As dataset sizes increase, and cloud computing costs drop, we are increasingly running the challenges in the cloud, swapping the paradigm from downloading data to uploading code. This can create an advantage that not only the code used for running the forward (trained model) is open and easy to inspect and use, but the framework used to train the model is also open for everyone to inspect. This removes the final barrier to code reuse.

However, there is another important opportunity here. With the increasing use of transfer learning, we are starting to see important gains in using large public datasets to pre-train models. Similarly, we are increasingly observing PhysioNet users combining datasets to improve the effectiveness of complex ML (and DL in particular). The rigorous modern paradigm for doing this is `Transfer Learning', where a ML model developed for one task is reused as the starting point for a model on a second task. Classic image networks such as VGGNet for face recognition have attracted much attention as starting points for thousands of new pieces of research \cite{simonyan2014very}. The challenges, and PhysioNet in general, provide a unique opportunity to create a repository of the equivalent DL networks for physiological signal processing and classification, in much the same way as ModelZoo.co and TFHub.dev have done for images and other related areas. This commons of models would lead to significant acceleration of the field and would allow groups to leverage private data of others, which the owners are not allowed to share (for various logistical, political, or other reasons). By posting a model trained on medical data from millions of in-house patients, another group could use such a model as a starting point and continue the training. If the new data set is only small, then the new model could be specific to that new population, but be less likely to overfit.

\section{Crowd sourcing and aggregation of weak and strong labels}
\label{sect:crowdsource}

Any public competition should produce several key products, including a collection of publicly accessible data, labels/predictions, algorithms/models, and publications that result from the event, which may lead to inventions, products, and other resources. In its most general sense, this is crowdsourcing. More specifically, the labels or predictions provided by each annotator can be aggregated to produce a more robust label on the data or resultant prediction. In the PhysioNet Challenges we leverage the inter-competitor agreement levels (measured by Fleiss' Kappa, $\kappa_F$) during the first (unofficial) 10 weeks of the competition. 
During the immediate following (week-long hiatus) period we identify the patients or events for which the prediction was most problematic. 
The assumption is that since the training and testing examples for which the $\kappa_F$ is the lowest are the hardest for most to classify, and therefore are most likely to be poorly labelled. 
These poorly labelled examples can then be reviewed in detail, and relabelled by multiple experts to raise the 
$\kappa_F$ towards that of the less equivocal examples. 
Of course, such an approach can be repeated over and over, depending on the time and resources available.
As long as there is no collaboration on labels (i.e. groupthink where the dominant authority in a group enforces their opinion) and the labels are independent, the labels should asymptotically approach some sort of `ground truth'.
This may be one of the most important contributions of any competition - the relabelling of the database to boost the accuracy of the labels. Without high confidence labels, there is little chance an automated algorithm, whether it is the latest ML algorithm, or just a simple threshold, will learn a set of weights (thresholds) that can label or predict accurately.

\begin{figure*}
\centering
\includegraphics[width=8cm]{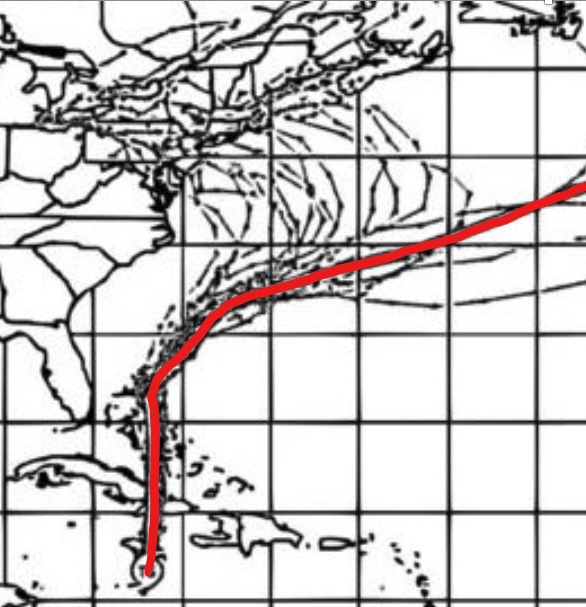}
\includegraphics[width=8cm]{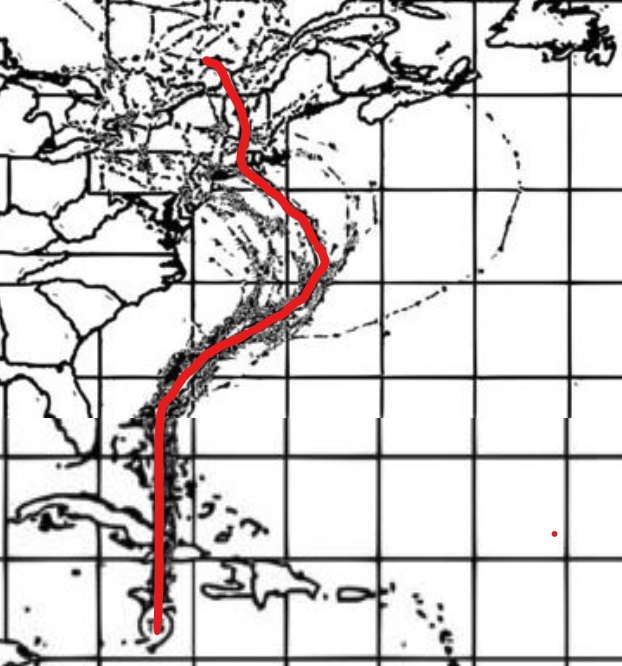}
\includegraphics[width=8cm]{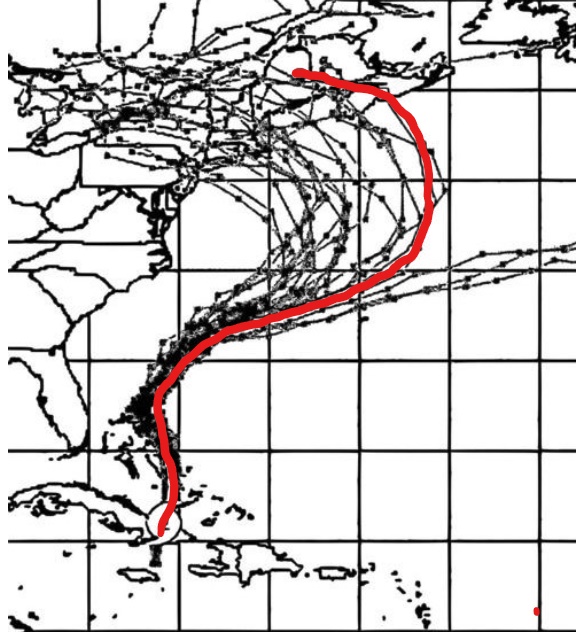}
\includegraphics[width=8.2cm]{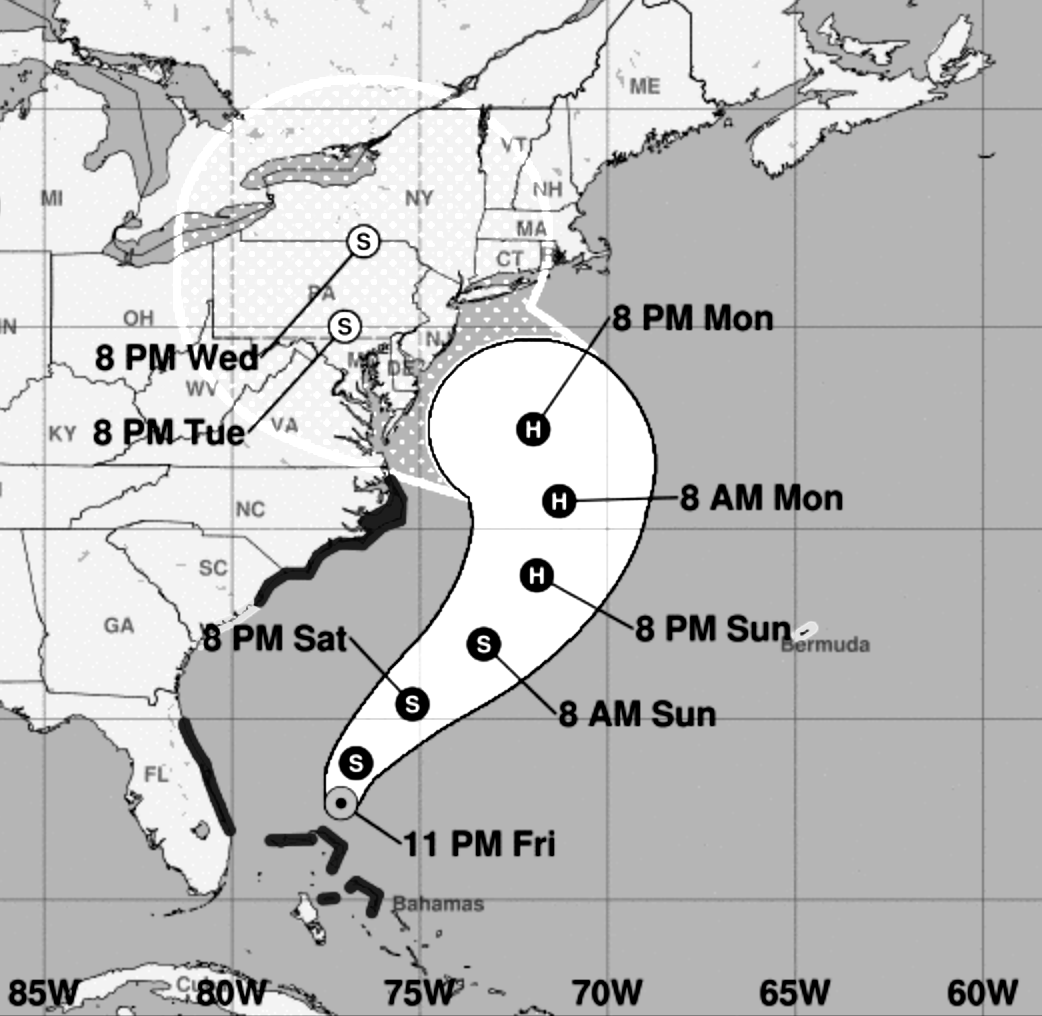}
\caption{\label{fig:Sandy} Hurricane tracks for various models of hurricane Sandy in 2012 as it passes over Jamaica and travels up the East coast of the USA. Red line indicates most probably path. Lower right image indicates cone of uncertainty. Figure adapted from \cite{NHC} under Creative Commons Attribution License 3.0 (CCAL).}
\end{figure*}

\section{Combining algorithms}
\label{sect:crowdsource2}

There is one field in which combining algorithms to deal with uncertainty is common: tropical cyclone (typhoon, hurricane, tropical storm or tropical depression) track prediction. For example, the Florida State Super Ensemble involves
11 models which are combined using a regression model, and produces forecasts better than each individual models or the mean of their predictions \cite{palmer2006predictability}.
The ever-increasing frequency of destructive weather events has made us all far more familiar with these ensemble models, and the fact that we tend to respond to landfall predictions when the models begin to converge. In other words, when models agree, our confidence in their ensemble prediction goes up, and we feel confident we need to act. Figure \ref{fig:Sandy} illustrates these storm tracks for Hurricane Sandy in 2012 as it progresses over Jamaica and up the Eastern seaboard of the USA, finally making landfall in New Jersey. The sudden pivot to the east in the last 24 hours before this final landfall was well-predicted after the hurricane had passed over Jamaica, because the complexities of modeling over landmasses (i.e. the Caribbean islands) made the uncertainties too high to make a call on where the final landfall might be.

The confidence intervals in the predictions are portrayed as a cone of uncertainty.
Inside the cone is the forecast line that represents the probable track of the center of a the tropical cyclone
over time increments in a set of circles up to five days ahead of the current observation. The size of the cone is based on the historical official forecast errors over five years.
It is therefore not reasonable to ask why we don't apply this form of ensemble averaging (of strong predictors) to the medical domain? In fact, this is what we have been doing over the last few years.

In previous challenges, we have shown that formally voting multiple algorithms together can be guaranteed to produce a higher performing algorithm than any single algorithm, without knowing the actual performance of any given algorithm. In particular, in past challenges, we have shown that anywhere from 20 to 50 algorithms provided significant gains in applications as wide as ECG repolarization interval estimation (to identify adverse drug events), to arrhythmia alarm calls, and prediction of sepsis \cite{Zhu2014,Zhu2015a,clifford2017af,PhysioNet2019}. The key to identifying which algorithms lies in determining the relative weights of each algorithm based on cross validation and regularization on the training data. However, many algorithms do not provide independent information, and so should not be weighted equally, even when they exhibit high performance. In earlier work we showed that a Bayesian voting framework can be used to combining any number of algorithms together and produce an estimate superior to any of the contributing algorithms on new, unseen data \cite{Zhu2014} \cite{Zhu2015a}. The weights of each algorithm are found by regressing their performances on training data, and features from the source data.
In this way, we discover the strengths of each algorithm, and the specific circumstances in which each one excels, and in which each one under-performs.
The combination is then highly likely to use the right algorithms in each particular context. This, of course, can only happen if each algorithm performs in very different ways, and this requires a measure of independence between algorithms.

Another useful byproduct of voting algorithms together is the variance over all the contributing algorithms can provide an approximation of the confidence one might have in any prediction.
Strangely, no medical monitor currently gives you any sense of how confident it is in its information. (It might have an intermediate `unsure' or `orange-light' warning, but that's not particularly useful for making decisions.
Feeding back a continuously evolving high-resolution confidence interval back to the clinical team may allow them to identify spurious one-off events to ignore, and see when it is time to request new information or try something new.
For example, if a confidence interval is continually increasing over time, it suggests the metric you are being fed is becoming less useful/relevant, and you should perhaps try to make an observation that would add value when integrated into the metric. The algorithm could even suggest which parameter is most likely to collapse the confidence intervals!
As Micheal Jordan noted ``unless you are actually doing the full-scale engineering statistical analysis to provide some error bars and quantify the errors, it's gambling'' \cite{Gomes2014}.

It should be noted that non-trivial algorithmic methods for voting have been around for decades, including expectation-maximization \cite{DawidandSkene1979}, products of experts \cite{Hinton00trainingproducts}, and Bayesian frameworks that account for both bias and variance \cite{Zhu2015a}. Although this is a large and evolving field,  it has mostly focused on combining categorical/ordinal estimates, and either humans {\em or} algorithms (as two separate research fields). There has been little work on combining independent humans and algorithms (in a non-collaborative manner). Since humans and algorithms are likely to demonstrate significant differences in their error distributions is important to consider the multimodal distributions of each type of voter, accounting for large differences in biases and variances, as we have done in Zhu {\it et al.} \cite{Zhu2014,Zhu2015a}.
This can also be thought of as a way to address different schools of thought around best practices.
The question to ask then is, should clinicians collaborate on a diagnosis, or should we use a formal framework to maintain independence? Mathematically speaking, we preserve more information by preventing collaboration. This seems counter-intuitive to clinical teams, where the discussion can lead to a higher confidence in the decision. However, the consensus may sometimes be driven by the biggest personality, the most articulate speaker, or the most senior person in attendance.

\section{Consensus of experts or independence of thought?}

It is worth noting that, for the first time, in twenty years of running these challenges, we also introduced a hackathon at the end of the challenge, one day before the start of the conference at which the challengers met to describe and defend their approaches. Although I criticized short hackathons earlier in this article (and have done for many years\footnote{As noted in section \ref{sect:challengelength}, a research project of less than six months is unlikely to lead to a useful result, and will require a lot of short cuts resulting in over-use of public code, reducing novelty, independence, and robustness.}), holding a hackathon at the end of a eight-month competition meant that all entrants had the benefit of exposure to the challenge for a significant period of time. Many challengers leave their best work until the last minute and then wish for a second chance. The hackathon provided just that chance.  Moreover, it provided the opportunity for challengers to work on-site with the competition organizers and collaborate. This provided a direct test to see if collaborative teams could produce better entries than a voting strategy.
This was an attempt to address the question of whether a consensus of experts is preferable to a weighted voting of independent algorithms.
We found that a weighted voting system of the top 10-20 
algorithms provided a boost
in the performance on all the hidden data, and in particular on the completely sequestered test set C, which proved problematic for almost all algorithms.
Most importantly, the outputs are always better than any individual algorithm, or the best individual algorithm on any given data set.

Perhaps even more importantly, the voting algorithm also beat all teams from the hackathon. The take-home message from this, if it holds in future experiments is, that maintaining independence provides a practical boost in performance and that no one algorithm was best, but a weighted vote of each algorithm provided a significant improvement over any single algorithm.
I make this key point in a recent keynote talk given at the Society of Critical Care Medicine \cite{CliffordCCMtalk2020}.

\section{Incentives to organize and participate in public competitions}
\label{sect:incentives}

Before I conclude this rather lengthy editorial, I wanted to address just {\em why} people will (And should) participate in public competitions or challenges.
Participation in such events is rarely for prize money. In 20 years, we have never advertised the award amounts of our prizes and only been asked once (via email) how much the purse might be. This is partially because we change the amounts from year to year, and partially because we want people entering it for the scientific challenge. We have found, through emails and informal conversations, that challengers enter:
\begin{itemize}
\item To gain early access to new data;
\item To supplement their research (often forming part of a doctoral or masters thesis);
\item To gain a deeper insight into a specific problem;
\item To pit their talents against the world's best groups;
\item To be part of a close-knit community, network and have the chance to publish early;
\item For the kudos of winning (often to enhance career prospects); and
\item To be the first to publish (or patent) on a new problem.
\end{itemize}

It is harder to understand why someone might run a challenge.  There is certainly some altruistic motivation, believing that data and code must exist in the public commons. There is also an element of feeling that we have benefited from public data and challenges in the past, and therefore are obliged to run these events. You could also argue that the publications by the challenge organizers describing each challenge become instantly highly cited (gaining over 100 citations in the first year or two). However, each challenge takes several years to run, and it would take a decade to increase a H-index by just 10 points in this way. I would argue that this is a pretty poor return in terms of impact factor, and that your efforts might be better focused elsewhere. I personally have many more highly cited manuscripts that took less than six months to complete.

Perhaps one of the most important incentives is the ability to define an important topic in the field, and push forward a branch of research. From a personal perspective, I find the insights that can be derived by comparing 50-100 approaches to the same problem on the same data is unique. You see themes that are hard to extract from the literature, because of the heterogeneity of their data, metrics and quality of articles. Often, novel and powerful features or approaches emerge from several groups, indicating that they are not just random luck or false discoveries, but are really new potential solutions or research directions. Inventions in science are often co-discovered almost simultaneously, because true leaps in our innovation and discovery are a product of all the shoulders on which we stand. We see this in the way that Nobel prizes are often awarded to multiple independent researchers for the same idea.
Novel breakthroughs rarely come from one genius or a single team, but grow in multiple areas, informing each other, and sparking ideas. By holding challenges we accelerate this process, converging teams on an annual basis. These teams then exchange ideas and spark new ones. The winning themes or discoveries bubble to the surface quickly. The gamification of science, if you like, can be a positive thing.

\section{Should grant awardees be required to run public competitions and should we use this to revamp peer-review and the process of selecting new grant awardees?}
Over the last two decades, the NIH, NSF, Wellcome Trust, and many other funding bodies and a few journals have enacted requirements for grantees/authors to disseminate their data. While this is a laudable idea, it is rarely enforced, and grantees usually only pay lip service to the idea. There are many reasons for this, ranging from not letting your competition get ahead of you and not having enough resources to disseminate data properly. The former problem can be obviated by placing a 2-4 year embargo on the data, allowing the researcher ample time to followup on the research and gain the next tranche of funding. It is not hard to argue that after seven to ten years, it is time to share the data, and give the rest of the academic community (and the taxpayers who funded the work) an opportunity to create something novel from the raw resources. The PhysioNet databases are an obvious example of how the return on investment for the researchers to make data public can be significant in terms of reputation, citation rates, and community building. 

Of course, without explicit funding for the publication of the data (and the support to understand the data), this process will rarely happen in any meaningful sense. I estimate it takes around one to two full-time researchers an entire year to curate the average dataset, and publish manuals and code that facilitate the use of the data. For highly complex datasets, like MIMIC II \cite{Saeed2011}, it took closer to five full-time research engineer years (not including the time it took to assemble the data in the first place). Perhaps the solution is to mandate a minimum effort level allocated to a grant to be used the year after the grant would normally end. 

Given that the return on investment of standard peer-reviewed grants has come into question in recent years
Obviously, we would be overwhelmed by challenges if this extended beyond some key grants mechanisms, and may indeed be a mechanism for a specific funding agency or foundation.  
\cite{Brian2011}, perhaps the requirements should be extended to include the requirement to run a challenge in this final year of funding for all recipients of large awards (say, over \$10 million, including renewals). 
The funding required to run a challenge is perhaps double that of disseminating the data as a static database, but arguably creates ten times the impact in just one or two years. Moreover, many large grants\footnote{such as the National Center for Advancing Translational Sciences Clinical and Translational Science Awards Program} already award small subgrants mid-stream to allow other researchers outside the original applicants to work on the data. Wouldn't it be a better use of money to take these funds and use them to disseminate the data more widely and support access to them based on results, rather than proposals?  
The logical end-point is then to award moderately large follow-on grants to the highest scoring challenge teams, (as well as the original researchers, to support continued research on the data and act as a central repository for curating the new products of the data). Given the relatively weak correlation between peer-reviewed grant application scores and the productivity of the researchers \cite{a6c0983d44f445068e7ee47bf1f78550}, it seems that an alternative method of awarding grants should be piloted. Of course, I am not proposing we throw the standard grant peer-review system out completely, but rather that a small subset of data-rich grants should be targeted to pilot the idea of awarding grants based on challenge performance, rather than peer-opinion. Perhaps we can even do a controlled study to compare matched researchers to evaluate the two systems?

\section{Final thoughts}

Beyond the impressive collection of data, open-source algorithms on which others can build, and the peer-reviewed high-quality articles that result from a public competition, all of which accelerate the field, there is a more exciting opportunity that it generates. That is, the opportunity to perform research on the way in which independent teams solve problems. We have seen over and over that a voting system derived from scores of multiple experts (or algorithms) can outperform any single expert (or algorithm), and this points the way towards the future. It seems to suggest that the old paradigm of using a single algorithm to make predictions in medicine is doomed to underperform and make biased decisions. In a field where there is an ever-increasing awareness of bias, it seems more important than ever. Moreover, a group of algorithms provides estimates in the confidence of a prediction or classification, allows a clinician to understand whether they should exercise caution before reacting, and perhaps retest in the near future, or even make new measurements to reduce the confidence intervals. Such an approach will allow us to bootstrap the quality of data sets, which in turn will lead to an improvement in performance of the algorithms.  This may be our only hope for dealing with the noisy labelling issue on enormous noisy data sets.

\section*{Acknowledgements}

This work was funded by the Gordon and Betty Moore Foundation and by the National Institute of General Medical Sciences (NIGMS) and the National Institute of Biomedical Imaging and Bioengineering (NIBIB) under NIH grant number 2R01GM104987-09. The content and all opinions are entirely the author's own, and do not belong to the Gordon and Betty Moore Foundation or the NIH. Of course, the thoughts in this article don't arise in a vacuum. Rather, they are the product of many years of conversations with mentors, mentees and colleagues. It's impossible to list them all, but it's important to call out Ary Goldberger, George Moody and Roger Mark, who had the vision and founded PhysioNet and these challenges over a decade before the ML field started running similar public events. George, in particular, was the driving force behind the detail in these challenges - a polymath with a mission (and the skills) to bring open-source/open-access science as a standard, long before the rest of the scientific community understood its importance (and acted upon it). You can't under-estimate the work and diligence it takes to assemble and run these challenges. I was lucky enough to learn from the best! Thank you Ary, George and Roger!

Thanks also to the wonderful members of my research group (both current and past), who have contributed so much.
So many have put in so much hard work. Joachim Behar, Qiao Li, Erick Andres Perez, Matt Reyna and Salman Seyedi have consistently contributed over the last few years. Assistance for generating figures 2 and 3 in the manuscript was gratefully received from Alistair Johnson and Matt Reyna respectively. Alistair Johnson and Matt Reyna did most of the hard work involved in generating figures 2 and 3, respectively. Matt Reyna read the first draft of this and had many useful suggestions. He has also taken on the large burden of co-leading the challenges with me in the last couple of years. All the errors and views in this article are, however, my own.

\bibliographystyle{unsrt}
\bibliography{references}

\begin{thebibliography}{10}

\bibitem{united1962medical}
United States. Army~Medical Service, A.L. Ahnfeldt, R.S. Anderson, J.B. Coates,
  C.H. Goddard, W.S. Mullins, and United States. Surgeon-General's Office.
\newblock {\em The Medical Department of the United States Army in World War
  II.}
\newblock Office of the Surgeon General, Department of the Army, 1962.
\newblock Available online at: http://resource.nlm.nih.gov/14120390R.

\bibitem{Wigner1960}
Eugene Wigner.
\newblock {The Unreasonable Effectiveness of Mathematics in the Natural
  Sciences}.
\newblock {\em Communications in Pure and Applied Mathematics,}, 13(I), 1960.

\bibitem{wikipediaSelfDrivingCarFatalities}
Anonymous.
\newblock {\em {List of self-driving car fatalities - Wikipedia}}, Last revised
  15 Jul, 2020 (Accessed 18 Jul, 2020).
\newblock Available online at:
  https://en.wikipedia.org/wiki/List{\_}of{\_}self-driving{\_}car{\_}fatalities.

\bibitem{CDC2017}
CDC.
\newblock {\em {Faststats - Accidents or Unintentional Injuries}}, Last
  revised: 20 Jan 2017 (Accessed 12 Dec 2019).
\newblock Available online at:
  https://www.cdc.gov/nchs/fastats/accidental-injury.htm.

\bibitem{PinnacleCare2015}
PinnacleCare.
\newblock {\em {The Human Cost and Financial Impact of Misdiagnosis}}, Last
  revised: 6 Jul 2016 (Accessed 12 Dec 2019).
\newblock Available online at:
  https://www.pinnaclecare.com/forms/download/Human-Cost-Financial-Impact-Whitepaper.pdf.

\bibitem{Slawomirski2017}
Luke Slawomirski, Ane Auraaen, and Niek Klazinga.
\newblock {\em {The Economics Of Patient Safety: Strengthening a value-based
  approach to reducing patient harm at national level}}, Last revised: 1 Mar
  2017 (Accessed 12 Dec 2019).
\newblock Available online at:
  https://www.oecd.org/els/health-systems/The-economics-of-patient-safety-March-2017.pdf.

\bibitem{WorldHealthOrganization2015}
{World Health Organization}.
\newblock {Global status report on road safety}.
\newblock {\em Injury Prevention}, Last revised: 2018 (Accessed 12 Dec 2019).

\bibitem{Gawande2007}
A~Gawande.
\newblock {\em {Better: A Surgeon's Notes on Performance}}.
\newblock Picador, 1st edition, 2007.

\bibitem{Balogh2016}
Erin~P. Balogh, Bryan~T. Miller, and John~R. Ball.
\newblock {\em {Improving diagnosis in health care}}.
\newblock National Academies Press, Jan 2016.

\bibitem{Attia2019}
Zachi~I. Attia, Peter~A. Noseworthy, Francisco Lopez-Jimenez, Samuel~J.
  Asirvatham, Abhishek~J. Deshmukh, Bernard~J. Gersh, Rickey~E. Carter, Xiaoxi
  Yao, Alejandro~A. Rabinstein, Brad~J. Erickson, Suraj Kapa, and Paul~A.
  Friedman.
\newblock {An artificial intelligence-enabled ECG algorithm for the
  identification of patients with atrial fibrillation during sinus rhythm: a
  retrospective analysis of outcome prediction}.
\newblock {\em The Lancet}, 394(10201):861--867, Sep 2019.

\bibitem{Saeed2011}
Mohammed Saeed, Mauricio Villarroel, Andrew~T. Reisner, Gari Clifford, Li~Wei
  Lehman, George Moody, Thomas Heldt, Tin~H. Kyaw, Benjamin Moody, and Roger~G.
  Mark.
\newblock {Multiparameter Intelligent Monitoring in Intensive Care II: A
  public-access intensive care unit database}.
\newblock {\em Critical Care Medicine}, 39(5):952--960, 2011.

\bibitem{Aboukhalil2008}
Anton Aboukhalil, Larry Nielsen, Mohammed Saeed, Roger~G. Mark, and Gari~D.
  Clifford.
\newblock {Reducing false alarm rates for critical arrhythmias using the
  arterial blood pressure waveform}.
\newblock {\em Journal of Biomedical Informatics}, 41(3):442--451, 2008.

\bibitem{Hug_CCM2011}
Caleb~W Hug, Gari~D Clifford, and Andrew~T Reisner.
\newblock {Clinician blood pressure documentation of stable intensive care
  patients: an intelligent archiving agent has a higher association with future
  hypotension.}
\newblock {\em Critical Care Medicine}, 39(5):1006--14, May 2011.

\bibitem{CliffordCCMtalk2020}
Gari~D Clifford.
\newblock {\em The Future of AI in Critical Care: A Tsunami of Predictions or a
  Consensus of Opinions?}, 16 Feb 2020 (Accessed Jun 3, 2020).
\newblock Available online at: youtu.be/Tv9ADevax8.

\bibitem{clifford2017af}
Gari~D Clifford, Chengyu Liu, Benjamin Moody, H~Lehman Li-wei, Ikaro Silva,
  Qiao Li, AE~Johnson, and Roger~G Mark.
\newblock {AF Classification from a short single lead ECG recording: the
  PhysioNet/Computing in Cardiology Challenge 2017}.
\newblock In {\em 2017 Computing in Cardiology (CinC)}, pages 1--4. IEEE, 2017.

\bibitem{RePEc:tin:wpaper:20020119}
J.S. Cramer.
\newblock {The Origins of Logistic Regression}.
\newblock Tinbergen Institute Discussion Papers 02-119/4, Tinbergen Institute,
  Last revised: Dec 2002 (Accessed 12 Dec 2019).
\newblock Available online at:
  https://ideas.repec.org/p/tin/wpaper/20020119.html.

\bibitem{NIPS2001_2020}
Andrew~Y. Ng and Michael~I. Jordan.
\newblock {On Discriminative vs. Generative Classifiers: A comparison of
  logistic regression and Na\"ive Bayes}.
\newblock In T.~G. Dietterich, S.~Becker, and Z.~Ghahramani, editors, {\em
  Advances in Neural Information Processing Systems 14}, pages 841--848. MIT
  Press, 2002.

\bibitem{NIPS2013_5073}
Nagarajan Natarajan, Inderjit~S Dhillon, Pradeep~K Ravikumar, and Ambuj Tewari.
\newblock Learning with noisy labels.
\newblock In C.~J.~C. Burges, L.~Bottou, M.~Welling, Z.~Ghahramani, and K.~Q.
  Weinberger, editors, {\em Advances in Neural Information Processing Systems
  26}, pages 1196--1204. Curran Associates, Inc., 2013.

\bibitem{Obermeyer447}
Ziad Obermeyer, Brian Powers, Christine Vogeli, and Sendhil Mullainathan.
\newblock Dissecting racial bias in an algorithm used to manage the health of
  populations.
\newblock {\em Science}, 366(6464):447--453, 2019.

\bibitem{PhysioNet2019}
Matthew~A. Reyna, Chris Josef, Russell Jeter, Supreeth~P. Shashikumar,
  M.~Brandon Westover, Shamim Nemati, Gari~D. Clifford, and Ashish Sharma.
\newblock Early prediction of sepsis from clinical data: the
  {P}hysio{N}et/{C}omputing in {C}ardiology {C}hallenge 2019.
\newblock {\em Critical Care Medicine}, 48:210--217, Feb 2020.

\bibitem{Clifford2011}
Gari~D. Clifford and Alistair E.~W. Johnson.
\newblock { Bed, death and beyond; prediction models and the ICU data
  superstore. (New Methods for Data Analysis in the ICU.)}.
\newblock In {\em World Conference on Computational Modeling of Cardiovascular
  and Cardiopulmonary Dynamics}, 2011.

\bibitem{Johnson2018-DBLP:journals/corr/abs-1812-02275}
Alistair E.~W. Johnson, Tom~J. Pollard, and Tristan Naumann.
\newblock Generalizability of predictive models for intensive care unit
  patients.
\newblock {\em arXiv Preprint Server}, abs/1812.02275, Last revised: 6 Dec 2018
  (Accessed 12 Dec 2019).
\newblock Available online at http://arxiv.org/abs/1812.02275.

\bibitem{Bender2016}
Eric Bender.
\newblock {Challenges: Crowdsourced solutions}.
\newblock {\em Nature}, 533:S62--S64, May 2016.

\bibitem{Ledford2017}
Heidi Ledford.
\newblock {Open-data contest unearths scientific gems-and controversy}.
\newblock {\em Nature}, 543(7645):299, Mar 2017.

\bibitem{Guinney2018}
Justin Guinney and Julio Saez-Rodriguez.
\newblock {Alternative models for sharing confidential biomedical data}.
\newblock {\em Nature Biotechnology}, 36(5):391--392, May 2018.

\bibitem{KAggleTermsofUse2019}
Kaggle.
\newblock {\em {Terms of Use | Kaggle}}, Last updated 20 Dec 2019 (Accessed 30
  Dec, 2019).
\newblock Available online at: https://www.kaggle.com/terms.

\bibitem{Blondel2012}
Vincent~D. Blondel, Markus Esch, Connie Chan, Fabrice Clerot, Pierre Deville,
  Etienne Huens, Fr{\'{e}}d{\'{e}}ric Morlot, Zbigniew Smoreda, and Cezary
  Ziemlicki.
\newblock {Data for Development: the D4D Challenge on Mobile Phone Data}.
\newblock {\em arXiv Preprint Server}, abs/1210.0137, sep Last revised: 28 Jan
  2013 (Accessed 12 Dec 2019).
\newblock Available online at http://arxiv.org/abs/1210.0137.

\bibitem{DeMontjoye2014}
Yves-Alexandre de~Montjoye, Zbigniew Smoreda, Romain Trinquart, Cezary
  Ziemlicki, and Vincent~D. Blondel.
\newblock {D4D-Senegal: The Second Mobile Phone Data for Development
  Challenge}.
\newblock {\em arXiv Preprint Server}, abs/1407.4885, Last revised: 30 Jul 2014
  (Accessed 12 Dec 2019).
\newblock Available online at http://arxiv.org/abs/1407.4885.

\bibitem{DREAMChallengesFAQs}
Dialogue for Reverse Engineering~Assessments and Methods.
\newblock {\em {FAQS - DREAM Challenges}}, 9 Oct, 2014 (Accessed 30 Dec, 2019).
\newblock Available online at: http://dreamchallenges.org/faqs/.

\bibitem{DREAMChallengesAbout}
Dialogue for Reverse Engineering~Assessments and Methods.
\newblock {\em {ABOUT DREAM - DREAM Challenges}}, 1 Oct, 2014 (Accessed 30 Dec,
  2019).
\newblock Available online at: http://dreamchallenges.org/about-dream/.

\bibitem{Andrae2015}
Anders Andrae and Tomas Edler.
\newblock On global electricity usage of communication technology: Trends to
  2030.
\newblock {\em Challenges}, 6(1):117–157, Apr 2015.

\bibitem{Andrae2017}
Anders Andrae.
\newblock {\em Total Consumer Power Consumption Forecast}, 10 2017.
\newblock Online at
  https://www.researchgate.net/publication/320225452{\_}Total{\_}Consumer{\_}Power{\_}Consumption{\_}Forecast.

\bibitem{DBLP:journals/corr/abs-1906-02243}
Emma Strubell, Ananya Ganesh, and Andrew McCallum.
\newblock Energy and policy considerations for deep learning in {NLP}.
\newblock {\em CoRR}, abs/1906.02243, 2019.

\bibitem{DBLP:journals/corr/abs-1803-03635}
Jonathan Frankle and Michael Carbin.
\newblock The lottery ticket hypothesis: Training pruned neural networks.
\newblock {\em CoRR}, abs/1803.03635, 2018.

\bibitem{ONeil2016}
Cathy O’Neil.
\newblock {\em Weapons of Math Destruction: How Big Data Increases Inequality
  and Threatens Democracy}.
\newblock Crown Publishers, 2016.
\newblock New York.

\bibitem{pmlr-v81-buolamwini18a}
Joy Buolamwini and Timnit Gebru.
\newblock Gender shades: Intersectional accuracy disparities in commercial
  gender classification.
\newblock In Sorelle~A. Friedler and Christo Wilson, editors, {\em Proceedings
  of the 1st Conference on Fairness, Accountability and Transparency},
  volume~81 of {\em Proceedings of Machine Learning Research}, pages 77--91,
  New York, NY, USA, 23--24 Feb 2018. PMLR.

\bibitem{10.1145/3306618.3314244}
Inioluwa~Deborah Raji and Joy Buolamwini.
\newblock Actionable auditing: Investigating the impact of publicly naming
  biased performance results of commercial ai products.
\newblock In {\em Proceedings of the 2019 AAAI/ACM Conference on AI, Ethics,
  and Society}, AIES ’19, page 429–435, New York, NY, USA, 2019.
  Association for Computing Machinery.

\bibitem{DBLP:journals/corr/abs-1710-03184}
Pratik Gajane.
\newblock On formalizing fairness in prediction with machine learning.
\newblock {\em CoRR}, abs/1710.03184, 2017.

\bibitem{DBLP:journals/corr/abs-1710-06876}
Samiulla Shaikh, Harit Vishwakarma, Sameep Mehta, Kush~R. Varshney,
  Karthikeyan~Natesan Ramamurthy, and Dennis Wei.
\newblock An end-to-end machine learning pipeline that ensures fairness
  policies.
\newblock {\em CoRR}, abs/1710.06876, 2017.

\bibitem{NIPS2017_6988}
Flavio Calmon, Dennis Wei, Bhanukiran Vinzamuri, Karthikeyan
  Natesan~Ramamurthy, and Kush~R Varshney.
\newblock Optimized pre-processing for discrimination prevention.
\newblock In I.~Guyon, U.~V. Luxburg, S.~Bengio, H.~Wallach, R.~Fergus,
  S.~Vishwanathan, and R.~Garnett, editors, {\em Advances in Neural Information
  Processing Systems 30}, pages 3992--4001. Curran Associates, Inc., 2017.

\bibitem{DBLP:journals/corr/abs-1808-00089}
Biplav Srivastava and Francesca Rossi.
\newblock Towards composable bias rating of {AI} services.
\newblock {\em CoRR}, abs/1808.00089, 2018.

\bibitem{Sahil2018}
Sahil Verma and Julia Rubin.
\newblock Fairness definitions explained.
\newblock {\em {ACM/IEEE International Workshop on Software Fairness}}, pages
  1--7, May 2018.

\bibitem{Bellamy2019}
Rachel Bellamy, Kuntal Dey, Michael Hind, Samuel Hoffman, Stephanie Houde,
  Kalapriya Kannan, Pranay Lohia, Jacquelyn Martino, Sameep Mehta, Aleksandra
  Mojsilovic, Seema Nagar, Karthikeyan Ramamurthy, John Richards, Diptikalyan
  Saha, Prasanna Sattigeri, Moninder Singh, Kush Varshney, and Yunfeng Zhang.
\newblock Ai fairness 360: An extensible toolkit for detecting and mitigating
  algorithmic bias.
\newblock {\em IBM Journal of Research and Development}, PP:1--1, 09 2019.

\bibitem{simonyan2014very}
Karen Simonyan and Andrew Zisserman.
\newblock Very deep convolutional networks for large-scale image recognition.
\newblock {\em arXiv preprint arXiv:1409.1556}, 2014.

\bibitem{NHC}
NOAA.
\newblock {\em Definition of the NHC Track Forecast Cone, National Hurricane
  Center and Central Pacific Hurricane Center}, 2020 (Accessed Jun 3, 2020).
\newblock Available online at: https://www.nhc.noaa.gov/aboutcone.shtml.

\bibitem{palmer2006predictability}
T.~Palmer and R.~Hagedorn.
\newblock {\em Predictability of Weather and Climate}.
\newblock Cambridge University Press, 2006.

\bibitem{Zhu2014}
Tingting Zhu, Alistair E~W Johnson, Joachim Behar, and Gari~D. Clifford.
\newblock {Crowd-sourced annotation of ECG signals using contextual
  information}.
\newblock {\em Annals of Biomedical Engineering}, 42(4):871--884, Apr 2014.

\bibitem{Zhu2015a}
Tingting Zhu, Nic Dunkley, Joachim Behar, David~A Clifton, and Gari~D Clifford.
\newblock {Fusing Continuous-Valued Medical Labels Using a Bayesian Model.}
\newblock {\em Annals of Biomedical Engineering}, 43(12):2892--902, Dec 2015.

\bibitem{Gomes2014}
Lee Gomes.
\newblock {Machine-Learning Maestro Michael Jordan on the Delusions of Big Data
  and Other Huge Engineering Efforts}.
\newblock {\em IEEE Spectrum Magazine}, 20 Oct 2014.
\newblock Online at:
  https://spectrum.ieee.org/artificial-intelligence/machine-learning/machinelearning-maestro-michael-jordan-on-the-delusions-of-big-data-and-other-huge-engineering-efforts.

\bibitem{DawidandSkene1979}
A.~P. Dawid and A.~M. Skene.
\newblock {Maximum Likelihood Estimation of Observer Error-Rates Using the EM
  Algorithm}.
\newblock {\em Journal of the Royal Statistical Society. Series C (Applied
  Statistics)}, 28(1):20--28, 1979.

\bibitem{Hinton00trainingproducts}
Geoffrey Hinton.
\newblock Training products of experts by minimizing contrastive divergence.
\newblock {\em Neural Computation}, 14(8):1771--1800, 2002.

\bibitem{Brian2011}
Brian~A. Jacob and Lars Lefgren.
\newblock {The impact of research grant funding on scientific productivity}.
\newblock {\em Journal of Public Economics}, 95(9-10):1168--1177, October 2011.

\bibitem{a6c0983d44f445068e7ee47bf1f78550}
{Ferric C.} Fang, Anthony Bowen, and Arturo Casadevall.
\newblock {NIH peer review percentile scores are poorly predictive of grant
  productivity}.
\newblock {\em eLife}, 5(e13323), February 2016.

\end{thebibliography}
\end{document}